# Towards Linking Histological Changes to Liver Viscoelasticity: A Hybrid Analytical-Computational Micromechanics Approach

Haritya Shah and Murthy N. Guddati*
*North Carolina State University, Raleigh, NC 27695-7908*
*\* Corresponding Author. Email: mnguddat@ncsu.edu*

ABSTRACT

Motivated by elastography that utilizes tissue mechanical properties as biomarkers for liver disease, with the eventual objective of quantitatively linking histopathology and bulk mechanical properties, we develop a micromechanical modeling approach to capture the effects of fat and collagen deposition in the liver. Specifically, we utilize computational homogenization to convert the microstructural changes in hepatic lobule to the effective viscoelastic modulus of the liver tissue, i.e., predict the bulk material properties by analyzing the deformation of repeating unit cell. The lipid and collagen deposition is simulated with the help of ad hoc algorithms informed by histological observations. Collagen deposition is directly included in the computational model, while composite material theory is used to convert fat content to the microscopic mechanical properties, which in turn is included in the computational model. The results illustrate the model's ability to capture the effect of both fat and collagen deposition on the viscoelastic moduli and represents a step towards linking histopathological changes in the liver to its bulk mechanical properties, which can eventually provide insights for accurate diagnosis with elastography.

## 1. INTRODUCTION

Elastography, either based on ultrasound (US) or magnetic resonance imaging (MRI), measures the elastic modulus of tissue, and uses it as a biomarker for various diseases including in the liver, such as liver fibrosis (see e.g., [1-4]). The fundamental premise is that disease causes microstructural changes, which in turn affects macroscopic tissue elasticity. Over the past decade, there is increased attention to viscosity of the tissue as a biomarker, in addition to elasticity, opening up diagnostic opportunities, e.g., for metabolic dysfunction-associated steatotic liver disease (MASLD) and metabolic dysfunction-associated steatohepatitis (MASH) [5-9]. The paper is aimed at linking the microstructural changes to mechanical properties, specifically in the context of the liver. While the proposed framework could be applicable to other diseases, we guide the development focusing on the MASLD, MASH and fibrosis. Further, the current investigation is motivated by wave-based elastography with small amplitude vibrations in the linear regime.

The histopathological changes through the evolution of liver diseases are very complex. Some of these include, the effects of softening associated with fat infiltration as well as hardening associated with fibrosis, the effects of nonlinearity associated with confinement due to fat infiltration and inflammation [10-12], and poroelastic effects of interstitial pressure changes associated with altered microchannel flow in the sinusoids [13-15]. While some of these effects, e.g., effects of simple steatosis, can be captured through analytical models, such models will likely not be able to capture the interplay between the many complex phenomena. Motivated by this, this study presents the first step of developing a computational homogenization framework that captures the effects of lipid and collagen deposition. The resulting framework can then be gradually refined by adding various complexities after careful examination of histopathological and physiological processes, eventually resulting in a tool that can be used to quantitatively link histopathological changes to changes in mechanical properties. Thus, the focus of the current paper is to develop such framework and perform preliminary illustration of the framework's ability to capture the effects of steatosis and fibrosis.

While computational modeling to link microstructural properties to bulk mechanical properties of the liver has not been well explored, there are significant contributions in analytical and observation-based approaches to either quantitatively or qualitatively explore these links. Parker and Ormachea [16] used classical composite theory to quantify the effect of lipid deposition and validated with oil-in-gel phantoms. Histological examinations coupled with elastography measurements have shown that viscoelasticity is affected by inflammation as well [5, 17-20]. Finally, the correlation between fibrosis and liver stiffness has long been established where correlations between stiffness and fibrosis grades have been developed based on extensive research including *in vivo* studies (see e.g., [1-4]). The studies, if quantitative, are largely limited to exploring the effects of fibrosis and steatosis, with respect to overall collagen and fat content respectively. They cannot be extended to include the effect of microstructural patterns of fat and collagen deposition, which may be a confounding factor in such correlations (see [21], which also provides analytical rationale). Driven by this, and with the expectation that the model can be gradually improved in complexity and precision, we propose a preliminary computational micromechanical model to capture the effects of steatosis and fibrosis.

Note that the proposed model is different from many existing microscopic computational models of the liver. For example, Wang and Jiang investigated shear wave propagation in liver with microstructure by direct numerical simulation and did not focus on the intermediate step of understanding homogenized (visco)elastic properties [22]. Ebrahem et al. investigated coupling poroelastic models with modeling blood flow in the liver and is not directly applicable to elastography investigations [23]. Yoshizawa et al. utilized agent based modeling to simulate the progression of fibrosis and not so much the effects of fibrosis [24]. We instead focus on a relatively simpler question that is more relevant to elastography: what is the link between the fat and collagen contents and the bulk tissue viscoelastic properties?

To answer the above question, we treat the liver to be a perfectly periodic assembly of identical hepatic lobules and apply computational homogenization to estimate the viscoelastic modulus for a given pattern of fat and/or collagen deposition. We also develop ad hoc microstructure generation algorithms to create synthetic microstructure profiles that mimic visual observations from histological observations. We then combine these two steps to present preliminary results for viscoelastic modulus as a function of fat content and/or collagen proportionate area (CPA), duly noting that the presented results are from a framework that needs to be refined with data from real observations. Notwithstanding this, the framework is expected to serve as a stepping stone for such refinements.

The paper is organized as follows. After making some observations on liver structure and physiology, the Theory section summarizes the main approach and presents underlying governing equations along with the computational homogenization approach that can link micromechanical properties to bulk moduli. The Methods section contains the approaches to generate the liver microstructure with fat and/or collagen, as well as the details of application of computational homogenization. The Results and Discussion section contains preliminary results for steatosis, fibrosis as well as combined steatosis and fibrosis, along with some observations based on the results. Finally, the paper is concluded with some closing remarks.

## 2. THEORY

### 2.1 Liver Structure and Physiology

The mechanics and mechanobiology of the liver involve complex microstructure and physiological processes. Liver is composed of tree-like vasculature composed on large vessels (hepatic vein, hepatic arteries, portal vein and bile duct), branching multiple times to form smaller vessels (central vein and portal triad), which interact through the hepatic lobules. Given our focus on elastography, which is typically focused on regions away from large vessels, our current goal is to understand the deformation of homogeneous liver tissue made up of repeating hepatic lobules.

The hepatocytes, which form the bulk of liver parenchyma interact mechanically with the microchannel flow within the sinusoids, making the process poroelastic [23]. The process is expected to be nonlinear due to the hyperelastic behavior of the parenchyma [25] and non-Newtonian nature of the blood flow [26]. Liver diseases change the microstructure due to, e.g., lipid infiltration of hepatocytes, ballooning, inflammation, and collagen deposition, which in turn result in changes in the nonlinearity and poroelasticity. Fortunately, in the context of wave-based elastography, which is the motivation behind this work, the situation simplifies significantly (by wave-based elastography, we mean ultrasound shear wave elastography, SWE, and magnetic resonance elastography, MRE). The high-frequency nature of wave deformation reduces the effect of flow-deformation interaction associated with MRE and SWE waves, effectively rendering the process (visco)elastic. Further, the small amplitude nature of the waves does not cause nonlinearity within the wave cycles, facilitating the probing of linearized viscoelastic properties of the current state of the liver tissue. Thus, the estimation of micromechanical changes in the liver from elastography measurements requires only linear elastic analysis of liver tissue away from the large vessels.

While the linearized material properties within a hepatic lobule may not be constant, as a first step, we will assume that the material properties are constant within the lobule (for the healthy liver without any lipid or collagen deposition). Furthermore, we do not explicitly model the effects of inflammation and ballooning and include them in the overall stiffness of the liver tissue. At this time, we only focus on the explicit effects of fat and collagen deposition. Additionally, given that MRE typically uses 50-100 Hz frequency range and SWE focuses on the 100-300 Hz range, the wavelength is generally more than 3 mm. Given the typical lobule size of ~1 mm, while there may be some dispersion effects at the upper end of the frequency range (300 Hz), these effects are not expected to be significant, and intralobular wave scattering effects can be neglected as a first step. With these simplifications, the viscoelastic properties of the bulk tissue in the liver can be computed from elastostatic analysis of repeating lobular structure, modified by lipid and collagen deposition. This is the focus of the remainder of the paper.

### 2.2 Overall Approach

Histological observations indicate that fat is deposited randomly within hepatocytes roughly as spherical inclusions. Parker and Ormachea [16] utilized this observation and combined it with classical composite materials theory [27] to provide analytical relationships between fat content and viscoelastic modulus. They approximate steatotic liver as a homogeneous tissue with random spherical inclusions of fat. Fat deposition, though random at the scale of hepatocytes, follow noticeable histological patterns at the scale of hepatic lobule. To account for this, we introduce a two-scale model of homogenization. Changes in localized viscoelastic modulus, at the intra-lobular scale, are predicted using the analytical formulation in [16]. This results in a varying distribution of viscoelastic properties within the hepatic lobule, which is then injected into a computational homogenization framework to predict the material properties of bulk liver tissue at larger scale.

As an extension to studying the effects of combined steatosis and fibrosis that occurs sometimes e.g., in MASH, we introduce collagen deposition into the computational model. This conforms with the histological observations of MASH where repeated steatotic shocks to the liver tissue eventually results in the onset of fibrosis. Given that the collagen length scale is larger than the hepatocytes, we directly induce collagen into the computational model of the hepatic lobule (as an initial step, this is done in an ad hoc manner to mimic histological observations). After examining the effect of collagen using this model, it is combined with the two-scale model for steatosis to provide some preliminary observations on combined steatosis and fibrosis, with some implications related to elastography.

In what follows, we will first summarize computational homogenization, followed by modeling steatosis, fibrosis, and then their combined effect.

## 2.3 Computational Homogenization

The basic idea of homogenization is to obtain material properties as experienced at macro scale, where the spatial scale of deformation is larger than the scale of the periodic cell, which is the hepatic lobule in the current study. Conceptually, homogenization is performed by applying macroscopic unit strain and examining the resulting microscopic tractions that can be converted to (averaged) macroscopic stress. Given the near incompressibility of liver tissue at the microscale, the homogenized compressibility would also be near zero. We thus focus on obtaining the homogenized shear modulus. This process is schematically illustrated with Figure 1, while the underlying details of the microstructure generation and deformation analysis are explained in the remainder of the paper. Figure 1a shows a rectangular region of the original microstructure containing repeated liver lobules (actual livers have some disorder in their microstructure, but we postulate that the assumed perfect periodicity captures the main effects of histopathology on averaged viscoelastic moduli). Figure 1b shows the deformed shape under applied shear strain at global scale; only the global shear strain is specified, while locally the tissue deforms satisfying the equations governing viscoelastic deformation. The homogenized stress is computed by analyzing the deformed shape with a framework known as asymptotic expansion homogenization (AEH, [28]). In what follows, we summarize the underlying equations as well as AEH within the context of viscoelastic deformation.

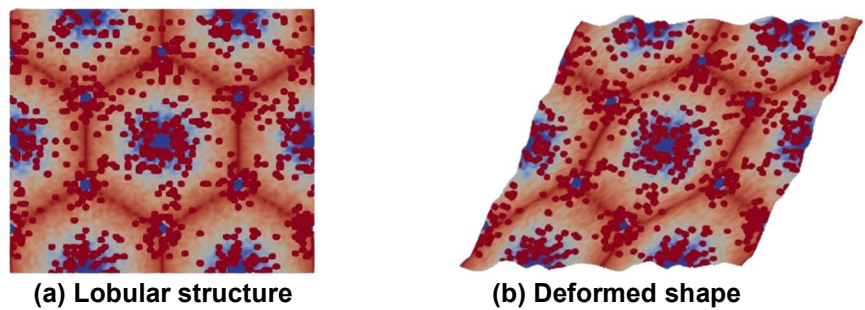

(a) Lobular structure     (b) Deformed shape

**Figure 1:** (a) Representative microstructure of the liver used in the current study (details of microstructure generation and analysis are presented later). The red spots depict average effects of collagen with high modulus, while the light blue regions represent softer tissue due to fat deposition near central veins. Dark blue regions represent central vein and portal triads. (b) deformed microstructure under imposed global shear strain, which not only contains the imposed global deformation but also the local oscillations in deformation due to microstructural heterogeneities

2.3.1 Viscoelastic Deformation: Governing Equations

The deformation of the liver tissue is governed by equilibrium equation, stress-strain and strain-displacement relations. Equilibrium equation is given by,

$$\nabla \cdot \boldsymbol{\sigma} = \mathbf{f} \tag{1}$$

where $\boldsymbol{\sigma}$ is the stress tensor, $\nabla$ is the spatial gradient ($\nabla \cdot$ is thus the divergence), and $\mathbf{f}$ is the body force vector. For a viscoelastic material, the stress $\boldsymbol{\sigma}$ is related to strain $\boldsymbol{\varepsilon}$ through the convolution integral:

$$\boldsymbol{\sigma} = \int_{-\infty}^{t} \mathbf{E}(t-\tau)\big(\boldsymbol{\varepsilon}(\tau) - \boldsymbol{\varepsilon}_0\big) d\tau, \tag{2}$$

where $t, \tau$ represent temporal variables and $\mathbf{E}$ is the relaxation modulus tensor. $\boldsymbol{\varepsilon}_0$ is the initial strain corresponding to the stress-free state. The strain is given by the symmetric gradient of the displacement $\mathbf{u}$:

$$\boldsymbol{\varepsilon} = \nabla_s \mathbf{u}. \tag{3}$$

General linear viscoelastic models for linear systems are most easily analyzed in the frequency domain, where the stress-strain relation in Equation (2) takes a much simpler product form:

$$\boldsymbol{\sigma}(\omega) = \mathbf{E}(\omega)\big(\boldsymbol{\varepsilon}(\omega) - \boldsymbol{\varepsilon}_0(\omega)\big), \tag{4}$$

where $\omega$ is the temporal frequency. Same notation is used for many variables in both time and frequency domains and the precise meaning is understood based on context. Assuming isotropy, the frequency-dependent complex modulus $\mathbf{E}(\omega)$ can be written in terms of complex-valued, frequency-dependent bulk and shear moduli. Given the near incompressibility of soft tissues, the deformation is dominated by shear distortion, governed only by the shear modulus, i.e., $G(\omega) = G'(\omega) + iG''(\omega)$, where $G', G''$ are storage and loss moduli respectively. Near incompressibility is simulated by taking the Poisson's ratio close to 0.5 (with appropriate selective reduced integration schemes when simulated with finite elements [29]). With such an approach, the elastic modulus tensor can be written as, $\mathbf{E}(\omega) = \mathbf{E}'(\omega) + i\mathbf{E}''(\omega)$, where the tensor $\mathbf{E}'$ captures the effect of the storage modulus and $\mathbf{E}''$ captures the effect of the loss modulus. We perform most of our computations at a frequency of 100 Hz, i.e. $\omega = 300\pi$ $rad/s$, at the intersection of MRE and SWE; sometimes the analysis is also performed at 300 Hz, to highlight some effects of the frequency on the histology-viscoelasticity relationships.

Complex arithmetic arising from complex valued $\mathbf{E}(\omega)$ is not available in most computational simulation software, including the one employed in our study (Multiphysics Object Oriented Simulation Environment, MOOSE [30]). To facilitate the current simulation in MOOSE, we convert the Equations 1,3,4 to real arithmetic separating real and imaginary parts of field and state variables (displacements, strains and stresses), i.e., $\mathbf{u} = \mathbf{u}_r + i\mathbf{u}_i$, $\boldsymbol{\varepsilon} = \boldsymbol{\varepsilon}_r + i\boldsymbol{\varepsilon}_i$, and $\boldsymbol{\sigma} = \boldsymbol{\sigma}_r + i\boldsymbol{\sigma}_i$. We similarly split the forcing function, $\mathbf{f} = \mathbf{f}_r + i\mathbf{f}_i$. Writing the variables in composite form, e.g., $\bar{\mathbf{u}} = \{\mathbf{u}_r, \mathbf{u}_i\}^T$, the equilibrium and strain-displacement relations in Equations 1,3 do not change in their form, except that the variables are replaced by the composite form of real and imaginary parts. On the other hand, the stress-strain relation, when expanded, takes the form:

$$\left\{\begin{array}{c}\boldsymbol{\sigma}_r \\ \boldsymbol{\sigma}_i\end{array}\right\} = \begin{bmatrix} \mathbf{E}' & \mathbf{E}'' \\ -\mathbf{E}'' & \mathbf{E}' \end{bmatrix} \left(\left\{\begin{array}{c}\boldsymbol{\varepsilon}_r \\ \boldsymbol{\varepsilon}_i\end{array}\right\} - \left\{\begin{array}{c}\boldsymbol{\varepsilon}_{0,r} \\ \boldsymbol{\varepsilon}_{0,i}\end{array}\right\}\right). \quad (5)$$

The above viewpoint indicates that the viscoelasticity equations in the frequency domain can be solved using real arithmetic with expanded form of field variables, coupled with the expanded form of the modulus tensor represented by the bracketed term in Equation 5. We implement this expanded form through MOOSE inputs, ultimately solving the final governing Equations 1,3,4.

### 2.3.2 Asymptotic Expansion Homogenization

Global (shear) strains on a periodic cell, needed for computational homogenization, can be applied with nonhomogeneous periodic boundary conditions. The resulting displacements can be processed for tractions on the boundary, which in turn can be averaged to obtain the global stresses. The end result would be global stress-strain relations and thus the effective modulus tensor. A mathematically equivalent technique that is more convenient to implement and is readily available in MOOSE is the asymptotic expansion homogenization algorithm (AEH, [28]). AEH separates the total displacement field to applied global displacement, and the local displacement which is the difference between the total displacement and the applied global displacement. The local displacement then satisfies simpler, homogeneous periodic boundary conditions. This perturbation is expanded in terms of Taylor series with respect to the size of the unit cell, and the low-order terms are equated to result in the effective modulus tensor. The final form of the effective homogenized modulus tensor $\mathbf{E}^H$ is:

$$\mathbf{E}^H_{ijkl} = \frac{1}{|Y|} \int_Y \mathbf{E}_{ijkl} \left( \mathbf{I} + \frac{\partial \mathbf{v}^{mn}_k}{\partial y_l} \right) dy, \quad (6)$$

where Einstein's indicial notation is used for convenience, with subscripts $i, j, k, l$ representing the coordinate directions. $Y$ is the unit cell, $\mathbf{E}$ is the modulus tensor that varies within $Y$. $\mathbf{v}^{mn}$ is the displacement variable from solving the governing equations with unit initial strain in $mn$ direction, i.e., the superscripts $m, n$ correspond to the two coordinate directions associated with the applied global strain. Given that only shear strain is applied in our setting, $m=1$ and $n=2$. Importantly, $\mathbf{v}^{mn}$ satisfies homogenous periodic boundary conditions:

$$\mathbf{v}^{mn}_{right} = \mathbf{v}^{mn}_{left}, \quad \mathbf{v}^{mn}_{top} = \mathbf{v}^{mn}_{bottom}, \quad (7)$$

where the self-explanatory subscripts refer to the boundaries of the periodic unit cell. The constructs of AEH including solving for $\mathbf{v}^{mn}$ and subsequent computation of $\mathbf{E}^H$ using Equation 6, are available in MOOSE, once the governing equation associated with Equation 5 is implemented. These constructs are utilized in the current work.

### 2.3.3 Unit Cell

A thin slice of liver tissue, when viewed under a microscope, will appear to have a repeating microstructure made up of hexagonal lobules with central vein at the center, and portal triads at each of the vertices. Though the primary repeating structure is hexagonal, given that AEH is derived and implemented in Cartesian coordinate system, it is simpler to utilize a rectangular unit cell to apply the AEH procedure. To this end, we propose to use a unit cell shown in Figure 2, made up of four halves of adjacent hexagons. The diameter of the hepatic lobule is assumed to be 1 mm in diameter based on histological observations. This translates to the rectangular repeating unit cell with a width of 0.86 mm and a height of 1.5 mm (by simple geometrical analysis). Note that the unit cell and all the analyses are two-dimensional. We do not consider three-dimensional volume because, (a) representing complex 3D packing of the lobules is complicated, (b) 3D simulations are computationally expensive, and (c) 3D packing of lobules is not completely well structured, indicating that the precise treatment of the complex behavior from (a) and (b) would not be of much value, if any.

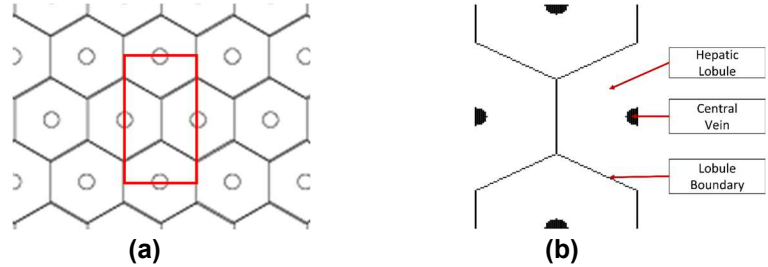

(a)  (b)

**Figure 2: (a) A 2D hexagonal lattice with the rectangular repeating cell highlighted in red and (b) layout of hepatic lobules in the identified rectangular unit cell.**

## 3. METHODS

### 3.1 Modeling Steatosis

#### 3.1.1 Intralobular fat and viscoelasticity distribution
To generate the distribution of viscoelastic modulus at the scale of multiple hepatocytes, we need to first generate spatial distribution of fat, mimicking the progression of steatosis in an actual liver. Since there has not been extensive quantitative analysis of fat distribution as a function of disease progression, we start with an ad hoc fat distribution function with the idea of revisiting this should it become an important factor. With such an approach, the fat percentage is assumed to be linearly varying with the square root of $x$, defined as the ratio of shortest distance from lobule boundary to the distance from the center of central vein. Thus, the fat fraction $V$ at any given spatial location is given by,

$$V = V_{out} + \left( \frac{\sqrt{x} - \sqrt{x_{min}}}{\sqrt{x_{max}} - \sqrt{x_{min}}} \right)\left( V_{in} - V_{out} \right), \qquad (8)$$

where $V_{in}$ is the fat fraction near the central vein while $V_{out}$ is the fat fraction near the lobule boundary.

Two patterns of fat distribution are considered. In Pattern 1, the fat content is the highest near the central vein, while in Pattern 2, the fat content is the highest near the boundary of the hepatic lobule. For both patterns, fat percentages are varied to control the total amount of fat present inside the hepatic lobule. The maximum fat percentage at any point within the lobule is capped at 50%, in line with the general histological observation of steatosis in the context of MASLD and MASH (see e.g., [31]), and consistent with the mathematical limitation arising from Christensen's theory of composite materials [27]. Lastly, given the random nature of fat deposition, 5% additive Gaussian noise was introduced to the fat distribution. Example fat distributions for both the patterns are shown in Figures 3a and 4a. The fat deposition pattern is computed on a rectangular grid of 80x160 points in the rectangular unit cell.

After populating the pixels in the domain of the unit cell with their respective fat concentration, the corresponding viscoelastic shear modulus is computed using analytical formulation using the formulas derived in [16]. This translates to a lobule with heterogeneous viscoelastic modulus distribution as shown in Figures 3b, 3c, 4b, and 4c. It is interesting to note that the loss modulus within the lobule does not vary as much as the storage modulus. This appears to be the algebraic outcome of the analytical formula, combined with the fact that the loss moduli of both fat and the tissue are smaller than the storage modulus of the tissue (at 100 Hz). The goal now is to simulate the deformation of the unit cell under prescribed macroscopic strain and use it to obtain the homogenized modulus, which is done using AEH summarized in Section 2.3.2 and finite element modeling.

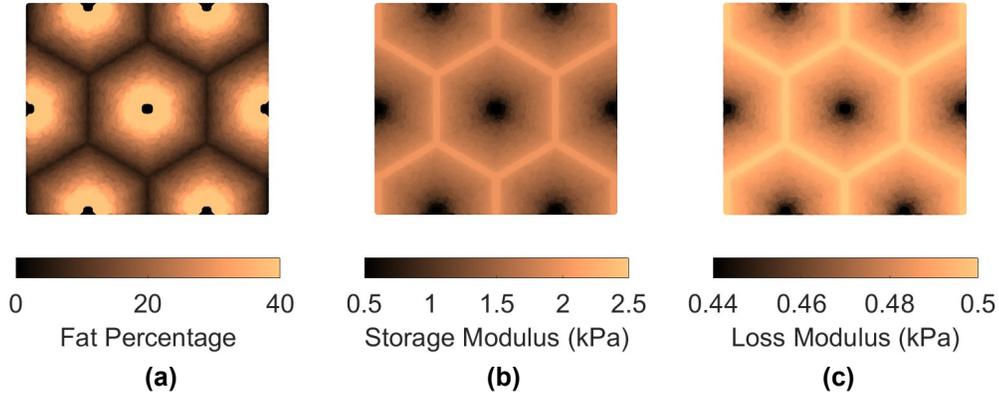

Figure 3: (a) fat distribution for Pattern 1 for average fat content of 16%; (b) resulting storage modulus distribution at 100 Hz; (c) loss modulus distribution at 100 Hz. Note that the ranges of storage and loss moduli are different and that the storage modulus varies significantly, while loss modulus varies by a much lesser extent.

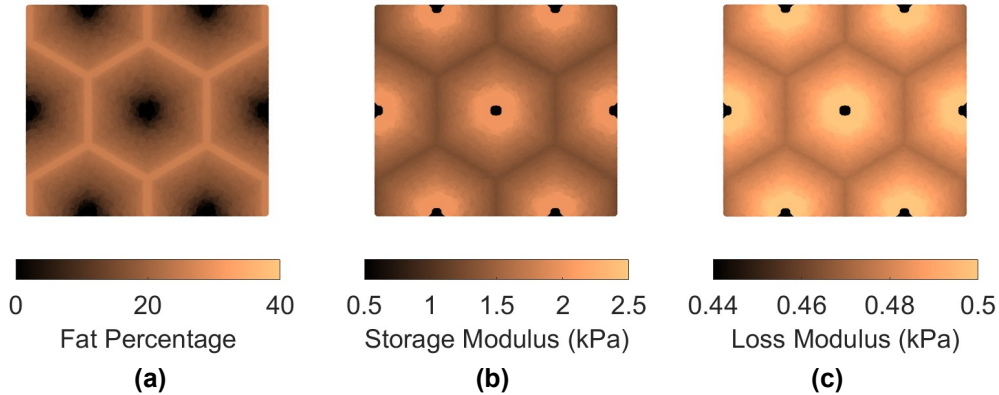

Figure 4: (a) fat distribution for Pattern 2 for average fat content of 16%; (b) resulting storage modulus distribution at 100 Hz; (c) loss modulus distribution at 100 Hz. Note that, similar to Pattern 1, the ranges of storage and loss moduli are different and that the storage modulus varies significantly, while loss modulus varies by a much lesser extent.

3.1.2 Finite Element Simulation

The unit cell in Figure 2 is discretized with 40x80 mesh of 8-node quadrilateral (QUAD8) finite elements, to capture the complicated displacement variation associated with heterogeneous structures. 2x2 Gauss quadrature was used for numerical integration. The viscoelastic modulus at the integration points are taken as the values at the closest point on the 80x160 grid used in the previous subsection. To verify that the accuracy of the chosen finite element mesh, convergence analysis was performed. For a fixed fat content (20%), mean homogenized modulus from 100 different realizations was used to calculate error.

The difference between the results from 40x80 and 80x160 meshes was found to be small (<0.5%). Hence, 40x80 elements mesh size was chosen for running the simulations.

## 3.2 Modeling Fibrosis

### 3.2.1 Distribution of Collagen

Based on its inherently unstructured progression, fibrosis is simulated with a probabilistic approach informed by general observations from histopathology. The probability of collagen deposition at any given point is defined as:

$$P = \Sigma\left(\frac{x}{d}\bigg|\mu,\sigma\right), \tag{9}$$

where $x$ is the distance from a particular location/region (defined based on the type of fibrosis, discussed later), $d$ is the control parameter that determines the spread of the deposition pattern and $\Sigma$ is Gaussian integral, i.e., the cumulative distribution function associated with the normal distribution with mean $\mu = 2$ and standard deviation $\sigma = 10$.

The definition of $x$ depends on the type of fibrosis. For periportal fibrosis, $x$ is taken as the distance to the closest portal triad, which has the effect of higher collagen deposition closer to the portal triad. For perisinusoidal fibrosis, $x$ is taken as the distance from the closest central vein. For bridging fibrosis, the $x$ is taken as the distance to the closest point to the edge where bridging occurs. If the bridging occurs between two portal triads, $x$ is the shortest (perpendicular) distance from the hexagonal edge. For bridging between central vein and portal triad, the shortest distance from lobular radii connecting the central vein to the portal triads is used.

Actual fibrosis is not isolated to perisinusoidal, periportal or bridging patterns but exists as a combination of these patterns, determined by CPA as well as the pattern of deposition. For stage 1 (F1, CPA≤5%), we assume that only perisinusoidal fibrosis exists. For stage 2 (F2, 5% < CPA ≤ 10%), the collagen deposition beyond 5% CPA follows periportal pattern. Fibrosis deposition beyond 10% is assumed to follow purely bridging pattern.

The control parameter $d$ is linearly interpolated based on CPA, with the interpolation limits chosen by trial and error, to visually match the deposition patterns found in the literature [32-34]. For perisinusoidal fibrosis, $d$ is interpolated from 0.75 mm at 1% CPA to 1.4 mm at 5% CPA. For periportal fibrosis, it is linearly interpolated form 0.75 mm at 5% CPA and 1.75 mm at 10% CPA. For bridging fibrosis, $d$ is linearly interpolated from 0.5 mm at 10% CPA to 1.0 mm at 20% CPA. The deposition pattern as well as the parameters used in the deposition algorithm are necessarily ad hoc and are driven by visual comparisons of subsequently obtained realizations to observations of patterns from literature [32-34]. It would be beneficial to make this more precise through quantitative analysis of existing histological observations, potentially complemented by simulations using mechanobiology [14]; these studies are outside the scope of the current effort. Nevertheless, the proposed deposition algorithm results in representative realizations of fibrosis at different CPAs as shown in Figure 5. As expected, the two patterns lead to the two different forms bridging fibrosis, starting at stage F3. Pattern 1 leads to bridging across the portal triads, while Pattern 2 leads to bridging between the portal triads and central vein.

Given the realistic nature of the generated fibrosis patterns, we utilize them in the micromechanical model for AEH to examine the changes in viscoelasticity with respect to changing CPA. Unlike in steatosis where fat concentration is converted into effective intralobular viscoelastic modulus, fibrosis does not need this step. Fat vesicles reside in hepatocytes but collagen deposition occurs at a larger scale around the hepatocytes and can thus be directly incorporated into the microstructural model.

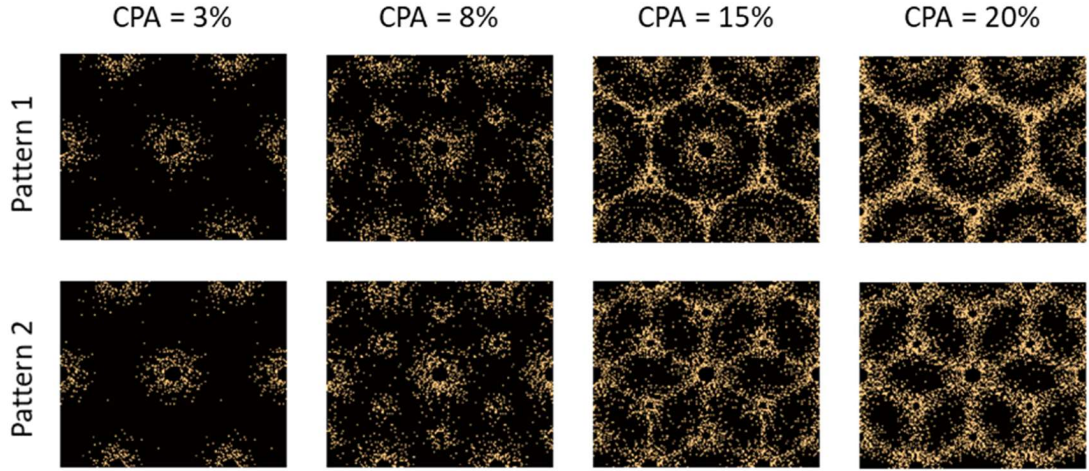

**Figure 5: Realizations of two different patterns of collagen distribution inside a hepatic lobule with increasing CPA. Early-stage fibrosis in both patterns is the same. The differentiating in Patterns 1 and 2 based on bridging fibrosis starts with a CPA of 10% and higher. The yellow dots within the realizations are collagen fibers.**

### 3.2.2 Finite Element Modeling

Unlike in steatosis where each point is assigned different viscoelastic moduli depending on the fat content, for fibrosis, each material point has a binary choice for modulus, either (viscoelastic) healthy liver tissue or (purely elastic) collagen. While such assignment appears simpler than steatosis on the outset, it leads to sharp changes in material properties, leading to stress concentrations that may not be representative of real physics. Fortunately, since our goal is not to understand local stresses but to estimate the changes in macroscopic moduli, we can significantly simplify the problem. We start with the same mesh as used for steatosis, i.e., with 40x80 rectangular elements, which may be understandably coarse compared to collagen deposition pattern. We perform finite element analysis with this mesh, where the integration points are assigned material properties based on the procedure described in the previous subsection. The rationale is that the implicit averaging performed through this approach would not lead to significant errors in macroscopic moduli which themselves are averaged quantities. Given that the homogenization is done over multiple realizations, we check this hypothesis by performing convergence analysis in a probabilistic sense. We essentially create 100 realizations of collagen deposition for 10% CPA and obtain the average viscoelastic modulus $4101 + 556i$ kPa. We repeat the process but with a refined mesh of 80x160 elements, which resulted in an average modulus of $4064 + 555i$ kPa. Given the small error of 0.89%, we conclude that 40x80 element mesh is sufficient for our current investigation.

### 3.3 Combined Modeling of Steatosis and Fibrosis

Steatosis and fibrosis can occur simultaneously in e.g., MASH, and it would be instrumental to examine the combined effects of CPA and fat content on both storage and loss moduli. This generalization of microstructure and associated finite element discretization is fairly straightforward: perform collagen deposition based on CPA, and for the remainder of the area, replace the healthy liver tissue by fatty tissue using algorithm in Section 3.1. Note that the average fat content used when applying the procedure in Section 3.1 needs to be adjusted by dividing with (1-CPA) as it is now the ratio of the volume of the fat to volume of the liver tissue not occupied by collagen. Once the intralobular fat distribution is determined, the points occupied by collagen take the corresponding elastic modulus (taken as 60 kPa in this study), while viscoelastic modulus is assigned at the remaining points based on the approach in Section 3.1. At the end, like in the case of steatosis and fibrosis, we end up with a unit cell with varying modulus which is analyzed using AEH and FEM to compute global viscoelastic modulus. To confirm the accuracy of 40x80 finite element mesh, we obtained average viscoelastic modulus for 15% fat and 10% CPA, with 40x80 mesh, and with 80x160 mesh. The difference in homogenized moduli from the two analyses is <2%. Thus the 40x80 mesh is deemed acceptable for simulating combined steatosis and fibrosis.

Advanced stage fibrosis (CPA > 10%) is accompanied by rapid loss of fat. Guided by this histopathological observation, we choose not to combine bridging part of collagen deposition with steatosis, i.e., CPA ranging from 1% to 10%, where Patterns 1 and 2 would be identical. Furthermore, given the prevalence of fat deposition in zone 3 [35-37], i.e. Pattern 1, we focus on this pattern with the average fat content varying from 0 to 30%.

## 4. RESULTS AND DISCUSSION

### 4.1 Effect of Steatosis

The procedure described in Section 3.1 is utilized to generate different realizations of fat content distributions on the unit cell, for varying overall fat content, for both Patterns 1 and 2. A total of 300 realizations are generated with fat content ranging from 0 to 40%, and AEH was applied at a forcing frequency of 100 Hz. For these simulations, The viscosity of fat at body temperature is taken as 0.4 Pa.s [16], translating to a complex modulus of G = 0.25i kPa at 100 Hz. Healthy liver tissue is assumed to have a shear modulus of 2 kPa and viscosity of 0.8 Pa.s, translating to complex modulus of (2.0+0.503i) kPa at 100 Hz. Figure 6a shows the variation of storage and loss moduli as a function of overall fat content. The deviation between the two patterns is most significant at the medium fat levels. In general, however, there does not appear to be much effect of the pattern of the fat deposition, compared to that of the overall fat content. Given this inconsequential nature of fat deposition pattern, the initial apparent ad hoc nature of the fat content generation algorithm in Equation 8 is deemed acceptable. Additionally, an apparent clinical implication is that viscoelasticity can potentially be used to quantify the fat content in the context of simple steatosis, after appropriate investigation of *in vivo* effects.

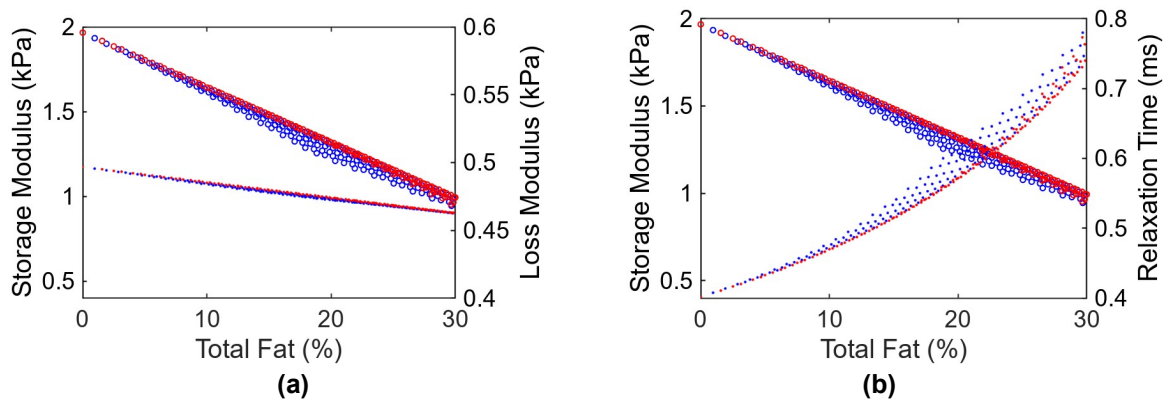

Figure 6: (a) plot of viscoelastic shear moduli as a function of % fat. Red and blue represent Patterns 1 and 2 in Figures 3 and 4. Circles represent storage moduli while dots represent loss moduli. (b) same data, but with loss modulus replaced by effective relaxation time, i.e. dots represent relaxation time.

As expected, the storage modulus decreases with increasing fat content, which is expected due to fluidity of fat droplets. This is consistent with *ex vivo* observations that indicate that simple steatosis leads to softening of the liver tissue [38]. On the other hand, *in vivo* observations indicate that even steatosis leads to increased stiffness of the liver [10, 11], which is attributed to the effects of confinement leading to increased stiffness. This can be captured with the help of hyperelastic models [25], but such generalization is outside the scope of the current effort.

Figure 6a indicates that not only the storage modulus, but the loss modulus decreases, albeit very slightly, with increasing fat content. While this may seem counterintuitive at the outset, it is reasonable in that the loss modulus of the fatty tissue (0.25 kPa) is less than the loss modulus of the healthy tissue (0.5 kPa). A more intuitive observation can be seen in Figure 6b where the results are plotted in terms of storage modulus and relaxation time $\tau = G'' / \omega G'$, i.e., the ratio of effective viscosity to elastic modulus. $\tau$ is essentially a measure of viscosity relative to elasticity, and follows the expected trend: fat deposition increases the relative viscosity.

## 4.2 Effect of Fibrosis

The procedure described in Section 3.2 is used to generate different realizations of collagen deposition inside the hepatic lobule, mimicking the progression of fibrosis across various stages. A total of 100 realizations with CPA ranging from 1% to 20% are generated. Similar to steatosis study in the previous section, the shear modulus of healthy tissue is taken as 2 kPa, and the viscosity is taken as 0.8 Pa.s. Given the modulus of the collagen is not well known, we consider various analyses with collagen shear modulus range identified from the literature [39-41], from 60 to 300 kPa, with an increment of 60 kPa. Figure 7a presents the composite storage and loss moduli as a function of CPA, for a collagen shear modulus of 60 kPa. Figure 7b presents the same data, as storage modulus and effective relaxation time. As expected, the storage modulus increases with increasing CPA, in contrast with the softening observed from fat deposition in Figure 6. On the other hand, the loss modulus slightly decreases with CPA, owing to the assumed absence of viscosity in the collagen. Consequently, the effective relaxation time, which is proportional to the ratio of loss modulus to storage modulus, decreases, as shown in Figure 7b. Note that the loss modulus in Figure 7a, at CPA ~ 0, is slightly below loss modulus of the matrix (0.5 kPa); this small discrepancy is attributed to the fact that the central vein region does not contribute to stiffness, which leads to reduction in stiffness, even the imaginary part.

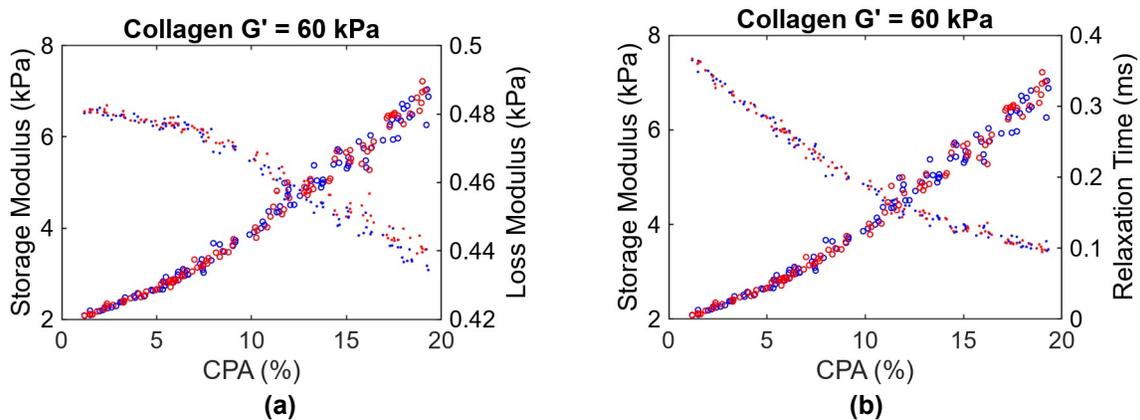

Figure 7: (a) plot of viscoelastic shear moduli as a function of CPA. Blue and red represent Patterns 1 and 2 in Figure 5. Circles represent storage moduli while dots represent loss moduli. (b) same data, but with loss modulus replaced by effective relaxation time.

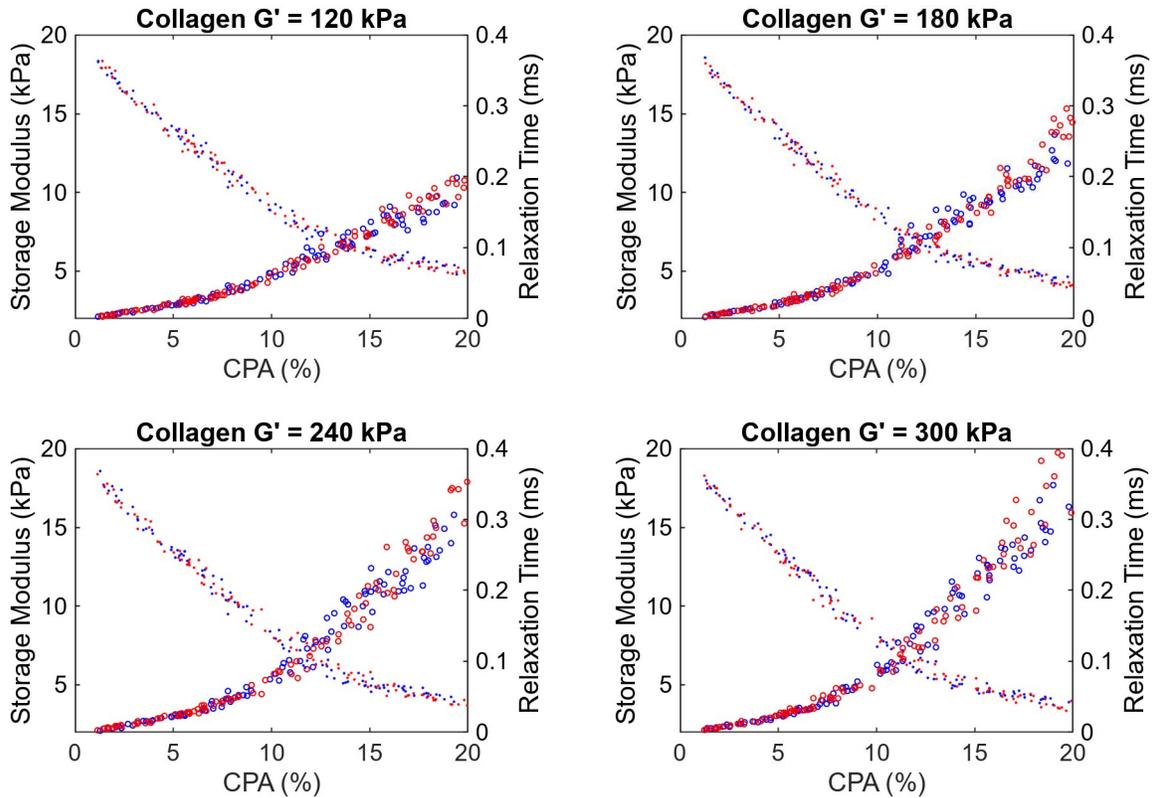

**Figure 8: Plots of changing storage modulus and relaxation time as a function of CPA, for different assumed shear modulus values of the collagen (120-300 kPa; 60 kPa results are shown in Figure 7b). Blue and red represent Patterns 1 and 2 in Figure 5. Circles represent storage moduli while dots represent loss moduli.**

Figure 8 provides storage modulus and relaxation time as a function of CPA, for the other four assumed collagen moduli, i.e. G'=120, 180, 240 and 270 kPa. It is evident that with an increasing CPA, the storage (shear) modulus increases and the relaxation time decreases. Moreover, for early fibrosis, up to 10% CPA, the scatter in the storage modulus as a function of CPA is not significant, indicating that there may be predictable relation between CPA and viscoelasticity parameters in the context of MASLD and MASH. Before using such relation with confidence, the collagen deposition patterns and modulus values should be determined more precisely and the results should be calibrated, potentially with animal models [31] to account for the *in vivo* effects not captured in the micromechanical model.

The plots in Figure 8 also show an accelerated increase in stiffening after CPA > 10%, especially for pattern 2 and for higher collagen modulus. This is attributed to the fact that the collagen deposition algorithm introduced a structure to the pattern for CPA > 10%. As observed in Figure 5, 15% CPA has a clear structure facilitating load paths through collagen rich regions, leading to increased effect of collagen on modulus increase. This is evident in Figure 5, with Pattern 2 forming a more efficient load paths compared to hexagonal load paths in Pattern 1; this is reflected in Figure 8, where the effective modulus for Pattern 2 is somewhat higher than that of Pattern 1 for CPA > 15%. Thus, in both cases, the stiffening is not just due to CPA, but also the structure of the collagen deposition pattern. Given this increased effect of the collagen, and given the random nature of its deposition, we also observe increased scatter of the predicted storage moduli for CPA > 10%. Note that significant quantitative conclusions should not be drawn from this study until careful investigation is performed into the collagen deposition patterns in real livers, as well as the representative modulus of the collagen in the liver.

## 4.3 Combined Effects of Steatosis and Fibrosis: Preliminary Observations

As a preliminary study to examine the effects of combined steatosis and fibrosis, we repeat the simulations, again at 100 Hz, for moderate CPA (ranging from 1 to 10%) and fat content ranging from 0 to 30%. Furthermore, we modify the healthy tissue properties to be power law [42], with a power of 0.15 and a coefficient of 800 (unit-consistent with Pa). The shear modulus of the collagen is taken as 60 kPa. The viscosity of fat is the same as before, 0.4 Pa.s.

The contour plots for storage and loss moduli are presented in Figures 9a and 9b, with a cross section of the contours presented in Figure 10a, where storage and loss moduli are presented as a function of percent fat, for two narrow CPA ranges of 1.9-2.1% and 7.9-8.1%. As expected the storage modulus reduces with increasing fat, and increases with increasing CPA. The loss modulus does not change much; this is not unexpected with respect to CPA, as the collagen is not viscous and affects only the effective storage modulus and not loss modulus. Interestingly, the loss modulus does not change (much) even with changing fat content. Closer examination reveals that at 100 Hz, the loss modulus of liver tissue (0.49i kPa) is quite close to the loss modulus of the fat (0.5i kPa), leading to this phenomenon. Triggered by this, we repeated the simulation at the higher frequency of 300 Hz, which can be considered as the upper end of ultrasound SWE. The results are presented in Figures 9c, 9d and 10b, which confirm that both storage and loss moduli change with CPA and fat content.

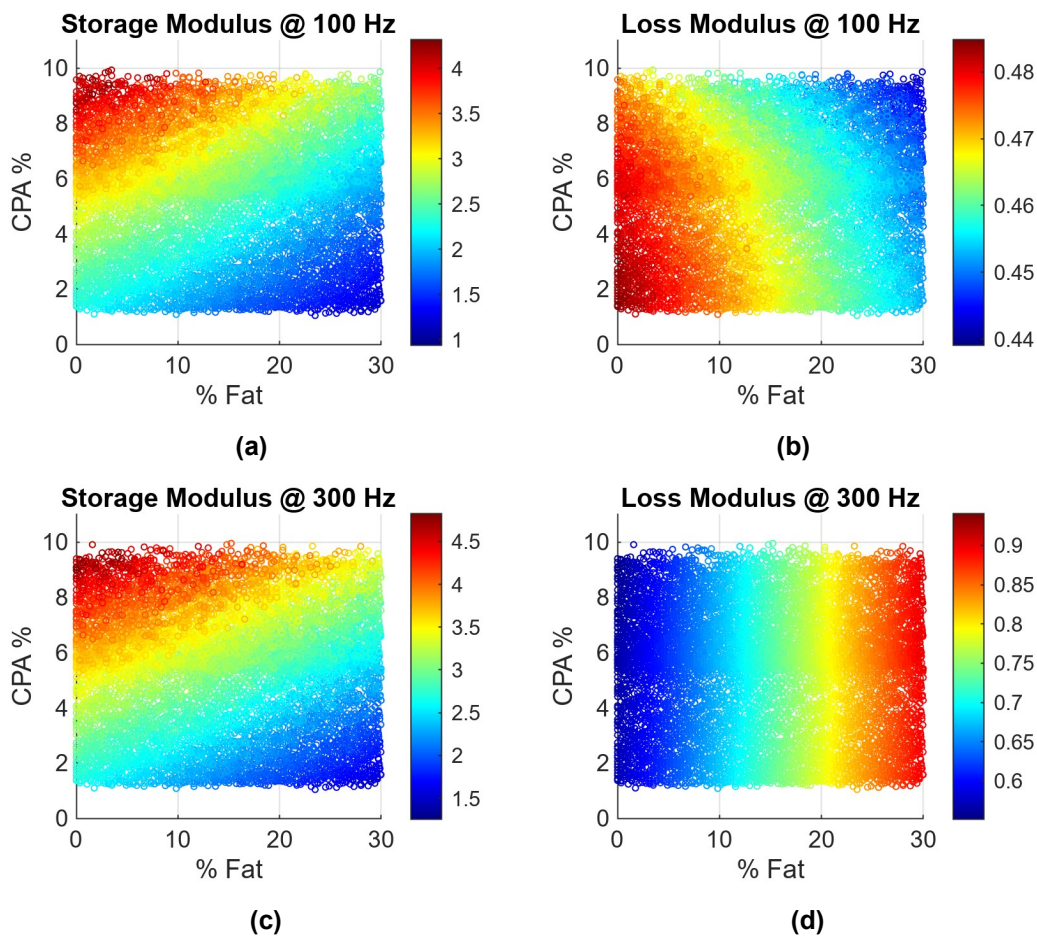

Figure 9: Variation of storage modulus (kPa) and loss modulus (kPa) as a function of fat fraction and CPA fraction. (a,b) represent storage and loss moduli for 100 Hz, while (c,d) represent the results for 300 Hz.

While the results in Figures 9 and 10 are preliminary, they can have important diagnostic implications. Looking at the specific case of Figure 9a, 9b, 10a at 100 Hz, given that only one quantity (storage modulus) is sensitive to fat content or CPA, it would not be possible to independently infer both CPA and fat content. On the other hand, using measurements at 300 Hz (Figure 9c, 9d and 10b), one can infer both CPA and fat content. While these models need to be revisited with careful inclusion of validated rheological models of liver tissue as well as elastic modulus of collagen in the liver, it appears that there is an important role frequency can play in estimating CPA and fat content from viscoelasticity measurements; ideally, it would be good to measure at multiple frequencies as done in ultrasound SWE and can be done in some versions of MRE.

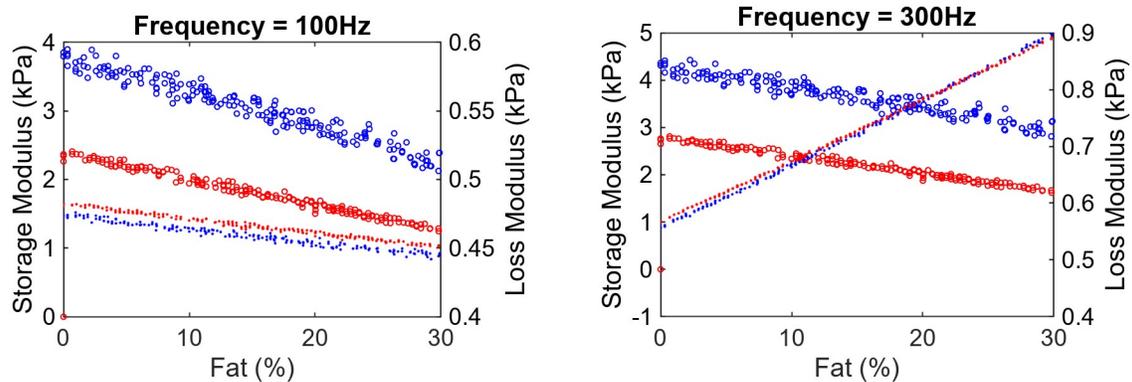

Figure 10: Cross-section of the contours in Figure 9. Storage modulus and loss modulus plotted as a variation of percent fat, for low CPA of 1.9=2.1 (red) and high CPA of 7.9=8.1 (blue). Circles represent storage modulus while dots represent loss modulus. Panel (a) is for 100 Hz, while (b) is for 300 Hz.

## 4. CONCLUSIONS

As a first step towards building a model that can link changes in the liver microstructure to its mechanical properties, we developed a computational modeling framework to capture the effects of fat and collagen deposition. The model is based on a computational homogenization approach that simulates the viscoelastic deformation of repeating rectangular unit cell of the liver to predict the macroscopic storage and loss modulus at different frequencies. The collagen is explicitly included in the computational model, while a composite-materials theory based analytical model is utilized to capture the effects of fat deposition. Ad hoc deposition patterns are utilized to visually mimic collagen and fat content distribution within the hepatic lobule, which are then analyzed with the proposed computational homogenization approach. It is found that relatively coarse finite element mesh can accurately compute the effective modulus, implying the practicality of the approach which requires multiple simulations given the random nature of the microstructure. We illustrated the framework to examine the effects of steatosis, fibrosis, and combined steatosis and fibrosis. Analysis of these preliminary results indicate that the mechanical changes due to steatosis are dependent on the total fat content, while the stiffening due to collagen is not only dependent on the overall content (CPA), but also the deposition pattern. Another interesting observation is that the frequency can play a role in relative sensitivity of CPA and fat content to the viscoelastic moduli. Before placing confidence in these studies, however, one must consider further enhancements described in Section 2.1.

As emphasized in multiple contexts, our study must be considered preliminary. While it can be used for making qualitative assessments, quantitatively linking viscoelasticity to fat content and CPA requires further investigations. Firstly, the deposition algorithms should be more informed by the histological data.

It would also be good to have a better handle on collagen modulus. Should the study be extended to interpreting higher-frequency SWE data with wavelengths comparable to lobule size, the homogenization should appropriately include inertial effects. Finally, *in vivo* effects need to be included, either due to confinement or due to changing interstitial pressure, which can perhaps be achieved through hyperelastic models coupled with poroelastic models. Model refinements could also include confinement effects from inflammation and ballooning on stiffness, and the resulting constricted flow in the sinusoids. Modeling such phenomena requires significant quantitative investigations into histopathology, micromechanics including the interaction between nonlinear flow and nonlinear deformation, along with uncertainty quantification. Finally, the developed models should be validated and/or calibrated with animal models to account for other *in vivo* effects that are not modeled explicitly. While we have not performed the studies listed in this paragraph, the proposed micromechanical model provides a practical computational framework for such systematic investigations, leading to insights into the links between disease states and mechanical properties measured through elastography.

## 5. ACKNOWLEDGEMENTS

The work is partially funded by National Science Foundation grant DMS-2111234, and by National Institute of Health grant R01 HL145268. The content is solely the responsibility of authors and does not necessarily represent the official views of the National Science Foundation, the National Heart, Lung, and Blood Institute, or the National Institutes of Health.

## 6. REFERENCES

1. Sandrin, L., et al., *Transient elastography: a new noninvasive method for assessment of hepatic fibrosis.* Ultrasound in Medicine & Biology, 2003. **29**(12): p. 1705-13.
2. Yin, M., K.J. Glaser, J.A. Talwalkar, J. Chen, A. Manduca, and R.L. Ehman, *Hepatic MR elastography: Clinical performance in a series of 1377 consecutive examinations.* Radiology, 2015. **278**(1): p. 114-124.
3. Ferraioli, G., et al., *WFUMB guidelines and recommendations for clinical use of ultrasound elastography: Part 3: Liver.* Ultrasound in Medicine & Biology, 2015. **41**(5): p. 1161-1179.
4. Muller, M., J.L. Gennisson, T. Deffieux, M. Tanter, and M. Fink, *Quantitative viscoelasticity mapping of human liver using supersonic shear imaging: preliminary in vivo feasability study.* Ultrasound in Medicine and Biology, 2009. **35**(2): p. 219-229.
5. Yin, M., et al., *Distinguishing between hepatic inflammation and fibrosis with MR elastography.* Radiology, 2017. **284**(3): p. 694-705.
6. Loomba, R., et al., *Novel 3D magnetic resonance elastography for the noninvasive diagnosis of advanced fibrosis in NAFLD: A prospective study.* Am J Gastroenterol, 2016. **111**(7): p. 986-994.
7. Yin, Z., et al., *Prediction of nonalcoholic fatty liver disease (NAFLD) activity score (NAS) with multiparametric hepatic magnetic resonance imaging and elastography.* European Radiology, 2019. **29**(11): p. 5823-5831.
8. Allen, A.M., et al., *Multiparametric magnetic resonance elastography improves the detection of NASH regression following bariatric surgery.* Hepatology Communications, 2020. **4**(2): p. 185-192.
9. Allen, A.M., et al., *The role of three-dimensional magnetic resonance elastography in the diagnosis of nonalcoholic steatohepatitis in obese patients undergoing bariatric surgery.* Hepatology, 2020. **71**(2): p. 510-521.
10. Liu, J., Y. Ma, P. Han, J. Wang, Y.-g. Liu, R.-f. Shi, and J. Li, *Hepatic steatosis leads to overestimation of liver stiffness measurement in both chronic hepatitis B and metabolic-associated fatty liver* *disease patients.* Clinics and Research in Hepatology and Gastroenterology, 2022. **46**(8): p. 101957.
11. Kimondo, J.J., R.R. Said, J. Wu, C. Tian, and Z. Wu, *Mechanical rheological model on the assessment of elasticity and viscosity in tissue inflammation: A systematic review.* Plos one, 2024. **19**(7): p. e0307113.
12. Shoham, N., P. Girshovitz, R. Katzengold, N.T. Shaked, D. Benayahu, and A. Gefen, *Adipocyte stiffness increases with accumulation of lipid droplets.* Biophysical journal, 2014. **106**(6): p. 1421-1431.
13. Caldwell, S., et al., *Hepatocellular ballooning in NASH.* Journal of hepatology, 2010. **53**(4): p. 719-723.
14. Mitten, E.K. and G. Baffy, *Mechanobiology in the development and progression of non-alcoholic fatty liver disease: an updated review.* Metabolism and Target Organ Damage, 2023. **3**(1): p. 2.
15. Parker, K.J., J. Ormachea, M.G. Drage, H. Kim, and Z. Hah, *The biomechanics of simple steatosis and steatohepatitis.* Physics in Medicine & Biology, 2018. **63**(10): p. 105013.
16. Parker, K.J. and J. Ormachea, *The quantification of liver fat from wave speed and attenuation.* Physics in Medicine & Biology, 2021. **66**(14): p. 145011-145011.
17. Zhang, X., et al., *Dynamic mechanical analysis to assess viscoelasticity of liver tissue in a rat model of nonalcoholic fatty liver disease.* Medical Engineering & Physics, 2017. **44**: p. 79-86.
18. RL, C.J.T.J.Y.M.G.K.S.S.E., *Early detection of nonalcoholic steatohepatitis in patients with nonalcoholic fatty liver disease by using MR elastography.* Radiology, 2011. **259**(3): p. 749-56.
19. Meyer, T., J. Castelein, J. Schattenfroh, A.S. Morr, R.V. da Silva, H. Tzschätzsch, R. Reiter, J. Guo, and I. Sack, *Magnetic resonance elastography in a nutshell: Tomographic imaging of soft tissue viscoelasticity for detecting and staging disease with a focus on inflammation.* Progress in Nuclear Magnetic Resonance Spectroscopy, 2024.
20. Sugimoto, K., et al., *US markers and necroinflammation, steatosis, and fibrosis in metabolic dysfunction–associated steatotic liver disease: the iLEAD study.* Radiology, 2024. **312**(2): p. e233377.
21. Poul, S.S. and K.J. Parker, *Fat and fibrosis as confounding cofactors in viscoelastic measurements of the liver.* Physics in Medicine & Biology, 2021. **66**(4): p. 045024.
22. Wang, Y. and J. Jiang, *Influence of tissue microstructure on shear wave speed measurements in plane shear wave elastography: a computational study in lossless fibrotic liver media.* Ultrasonic Imaging, 2018. **40**(1): p. 49-63.
23. Ebrahem, A., E. Jessen, M.F. ten Eikelder, T. Gangwar, M. Mika, and D. Schillinger, *Connecting continuum poroelasticity with discrete synthetic vascular trees for modelling liver tissue.* Proceedings of the Royal Society A, 2024. **480**(2285): p. 20230421.
24. Yoshizawa, M., M. Sugimoto, M. Tanaka, Y. Sakai, and M. Nishikawa, *Computational simulation of liver fibrosis dynamics.* Scientific reports, 2022. **12**(1): p. 14112.
25. Wex, C., S. Arndt, A. Stoll, C. Bruns, and Y. Kupriyanova, *Isotropic incompressible hyperelastic models for modelling the mechanical behaviour of biological tissues: a review.* Biomedical Engineering/Biomedizinische Technik, 2015. **60**(6): p. 577-592.
26. Rani, H., T.W. Sheu, T. Chang, and P. Liang, *Numerical investigation of non-Newtonian microcirculatory blood flow in hepatic lobule.* Journal of biomechanics, 2006. **39**(3): p. 551-563.
27. Christensen, R., *Viscoelastic properties of heterogeneous media.* Journal of the Mechanics and Physics of Solids, 1969. **17**(1): p. 23-41.
28. Hales, J., M. Tonks, K. Chockalingam, D. Perez, S. Novascone, B. Spencer, and R. Williamson, *Asymptotic expansion homogenization for multiscale nuclear fuel analysis.* Computational Materials Science, 2015. **99**: p. 290-297.

29. Hughes, T.J.R., *The Finite Element Method: Linear Static and Dynamic Finite Element Analysis*. 2012: Dover Publications. 1-704.
30. Giudicelli, G., et al., *3.0-MOOSE: Enabling massively parallel multiphysics simulations.* SoftwareX, 2024. **26**: p. 101690.
31. Khoury, M., et al., *Glycogen synthase kinase 3 activity enhances liver inflammation in MASH.* JHEP Reports, 2024. **6**(6): p. 101073.
32. Amzolini, A.M., et al., *Triglyceride and glucose index: a useful tool for non-alcoholic liver disease assessed by liver biopsy in patients with metabolic syndrome?* Romanian Journal of Morphology and Embryology, 2021. **62**(2): p. 475.
33. Hano, H., S. Takasaki, Y. Endo, T. Harada, K. Komine, and Y. Koike, *Histological reassessment of the role of bridging fibrosis in the angioarchitectural features associated with lobular distortion of the liver in chronic viral hepatitis.* Hepatology Research, 2016. **46**(3): p. E70-E78.
34. Hano, H., S. Takasaki, H. Kobayashi, T. Koyama, T. Lu, and K. Nagatsuma, *In the non-cirrhotic stage of nonalcoholic steatohepatitis, angioarchitecture of portal veins and lobular architecture are maintained.* Virchows Archiv, 2013. **462**: p. 533-540.
35. Brunt, E.M. and D.G. Tiniakos, *Histopathology of nonalcoholic fatty liver disease.* World journal of gastroenterology: WJG, 2010. **16**(42): p. 5286.
36. Chalasani, N., L. Wilson, D.E. Kleiner, O.W. Cummings, E.M. Brunt, A. Ünalp, and N.C.R. Network, *Relationship of steatosis grade and zonal location to histological features of steatohepatitis in adult patients with non-alcoholic fatty liver disease.* Journal of hepatology, 2008. **48**(5): p. 829-834.
37. Kleiner, D.E. and H.R. Makhlouf, *Histology of NAFLD and NASH in adults and children.* Clinics in liver disease, 2015. **20**(2): p. 293.
38. Li, D., P.A. Janmey, and R.G. Wells, *Local fat content determines global and local stiffness in livers with simple steatosis.* FASEB Bioadv, 2023. **5**(6): p. 251-261.
39. Wang, Y. and J. Jiang, *A two-dimensional (2d) systems biology-based discrete liver tissue model: A simulation study with implications for ultrasound elastography of liver fibrosis.* Computers in biology and medicine, 2019. **104**: p. 227-234.
40. Gautieri, A., S. Vesentini, A. Redaelli, and M.J. Buehler, *Hierarchical nanomechanics of collagen microfibrils.* Nature Precedings, 2010: p. 1-1.
41. Khamdaeng, T., J. Luo, J. Vappou, P. Terdtoon, and E. Konofagou, *Arterial stiffness identification of the human carotid artery using the stress–strain relationship in vivo.* Ultrasonics, 2012. **52**(3): p. 402-411.
42. Parker, K.J., *Power laws prevail in medical ultrasound.* Physics in Medicine & Biology, 2022. **67**(9): p. 09TR02.